\documentclass[journal=nalefd,manuscript=article]{achemso}

\usepackage[version=3]{mhchem} % Formula subscripts using \ce{}
\usepackage{amsmath,chemformula}

\author{John Brewer}
\affiliation[The University of California, Los Angeles]
{Department of Materials Science and Engineering, University of California, Los Angeles, Los Angeles, CA}
\author{Matthew F. Campbell}
\affiliation[University of Pennsylvania]
{Department of Mechanical Engineering and Applied Mechanics, University of Pennsylvania, Philadelphia, PA}
\author{Pawan Kumar}
\affiliation[University of Pennsylvania]
{Department of Electrical and Systems Engineering, University of Pennsylvania, Philadelphia, PA}
\author{Sachin Kulkarni}
\affiliation[The University of California, Los Angeles]
{Department of Materials Science and Engineering, University of California, Los Angeles, Los Angeles, CA}
\author{Deep Jariwala}
\affiliation[University of Pennsylvania]
{Department of Electrical and Systems Engineering, University of Pennsylvania, Philadelphia, PA}
\author{Igor Bargatin}
\affiliation[University of Pennsylvania]
{Department of Mechanical Engineering and Applied Mechanics, University of Pennsylvania, Philadelphia, PA}
\author{Aaswath P. Raman}
\affiliation[The University of California, Los Angeles]
{Department of Materials Science and Engineering, University of California, Los Angeles, Los Angeles, CA}
\email{aaswath@ucla.edu}

\title{Multi-scale photonic emissivity engineering for relativistic lightsail thermal regulation}

\abbreviations{}

\begin{document}

\begin{abstract}
The Breakthrough Starshot Initiative aims to send a gram-scale probe to Proxima Centuri B using a laser-accelerated lightsail traveling at relativistic speeds. Thermal management is a key lightsail design objective because of the intense laser powers required but has generally been considered secondary to accelerative performance. Here, we demonstrate nanophotonic photonic crystal slab reflectors composed of 2H-phase molybdenum disulfide and crystalline silicon nitride, highlight the inverse relationship between the thermal band extinction coefficient and the lightsail's maximum temperature, and examine the trade-off between the acceleration distance and setting realistic sail thermal limits, ultimately realizing a thermally endurable acceleration minimum distance of 16.3~Gm. We additionally demonstrate multi-scale photonic structures featuring thermal-wavelength-scale Mie resonant geometries, and characterize their broadband Mie resonance-driven emissivity enhancement and acceleration distance reduction. Our results highlight new possibilities in simultaneously controlling optical and thermal response over broad wavelength ranges in ultralight nanophotonic structures. 
\end{abstract}

Humanity's exploration of interstellar space has long been considered to be a topic for science fiction, rather than contemporary scientific inquiry. While exploration within our solar system can and has been accomplished by numerous propulsion schemes, the light year-scale distances involved in \textit{interstellar} travel pose fundamental scientific and engineering challenges that cannot as of yet be addressed by even advanced nuclear pulse \cite{etde_5083544} or fusion power-based \cite{beals1988project,Bond1978,LONG20111820} methods. Therefore, the Breakthrough Starshot Initiative has proposed an ambitious and potentially feasible means of achieving interstellar travel within human lifetimes, by using a laser-driven lightsail that is accelerated to one-fifth the speed of light to tow a small gram-scale probe 4.2 light years to Earth's nearest habitable candidate exoplanet, Proxima Centauri B\cite{Wagner2021,Dumusque2012,Parkin2018}. Lightsails are highly reflective surfaces that are propelled via photon momentum transfer\cite{novotny_hecht_2012}.
In the time since this phenomenon was first observed by Maxwell in 1873, lightsail-driven vehicles have been studied and theorized extensively for solar\cite{garwin1958solar,drexler1979high,Davoyan2021,Tsuda2011}, astrophysical\cite{Lingam_2020,doi:10.2514/6.2020-3537}, and  laser\cite{Forward1984,Kulkarni_2018,Redding1967,Kipping2017} sources. In the latter case, Starshot's target velocity and destination require using a gigawatt-scale irradiance phased-array laser source, resulting in enormous light intensities on the sail and creating a unique and difficult problem in sail design and material selection.

Realizing functional laser-driven sails necessitates solving a series of problems that lie at the intersection of material, mechanical, and optical engineering\cite{Atwater2018}, solutions to which have only recently been put within reach through advances in nanofabrication\cite{Chou1996,Chen2015,Johnson2014}, radiative cooling\cite{Raman2014,Rephaeli2013}, and photonics\cite{Kolodziejski1999,Fan2002,Molesky2018,Yu2014}. As the sail accelerates, incident laser light will become redshifted in the sail's frame of reference. This sets a restriction on usable sail film materials to those that have little or no measurable absorption over the entire redshifted laser band. The sail must furthermore be highly reflective over the laser band in order to accelerate to its ultimate speed in as short a distance as possible and thereby limit the laser-on time. In addition to reflectivity, the sail must also exhibit sufficient mechanical robustness to survive the extreme acceleration-induced forces, as well as its interaction with the interstellar medium\cite{Hoang_2017,Hoang_2017_2,doi:10.2514/2.3595}. While accelerating, the sail must be shaped\cite{campbell2021relativistic} or patterned properly so as to stably ride the laser beam when faced with non-ideal beam shapes and alignments\cite{Manchester_2017}. Finally, the sail must possess sufficient emissivity to effectively radiate heat generated due to any residual absorption of the sail material.

Motivated by the Breakthrough Starshot Initiative's goals, pioneering work on this topic proposed and analyzed a range of lightsail designs that minimize laser requirements\cite{Atwater2018}, suggested the use of highly thermally emissive material layers for radiative cooling, examined how the sail material's laser band absorptivity can affect sail temperatures, pointed to the need for realistic thermal limits, and demonstrated an accelerative trade-off between sail reflectivity and mass through a newly defined figure of merit\cite{Ilic2018}. This was followed by demonstrations of methods to passively stabilize beam-riding surfaces using spherical, parabolic, and conical sail shapes\cite{Ilic2019}. Passive beam-riding stability of curved lightsails was then adopted to design and analyze flat metasurface and diffractive beam stable structures and the opto-mechanical considerations of these sails, showing that their designs were able to meet a material melting point-based thermal sail limit\cite{Salary2020,Salary2021}. Recent work has proposed designs that achieve even lower acceleration distances using generalized gradient descent and topological optimization methods for nanophotonic sail design\cite{doi:10.1021/acsphotonics.0c00768}. While these works have demonstrated key advances in sail engineering, thus far they have only explored the use of Si, SiO\textsubscript{2}\cite{Ilic2018,Salary2020,Salary2021}, and Si\textsubscript{3}N\textsubscript{4}\cite{doi:10.1021/acsphotonics.0c00768,Gao2020} as sail materials. Additionally, the literature up to this point has not benchmarked photonic sail designs for acceleration distance under realistic thermal constraints, or determined furthermore how photonic designs might simultaneously enhance emissivity over infrared wavelengths while also minimizing the sail mass and maximizing its laser reflection. More broadly, given an increasingly large range of possible sail designs and materials, it is important to develop a methodology by which one can select sail designs constrained by their thermal performance and material degradation limits. 

\begin{figure}[h]
\includegraphics[width=16cm]{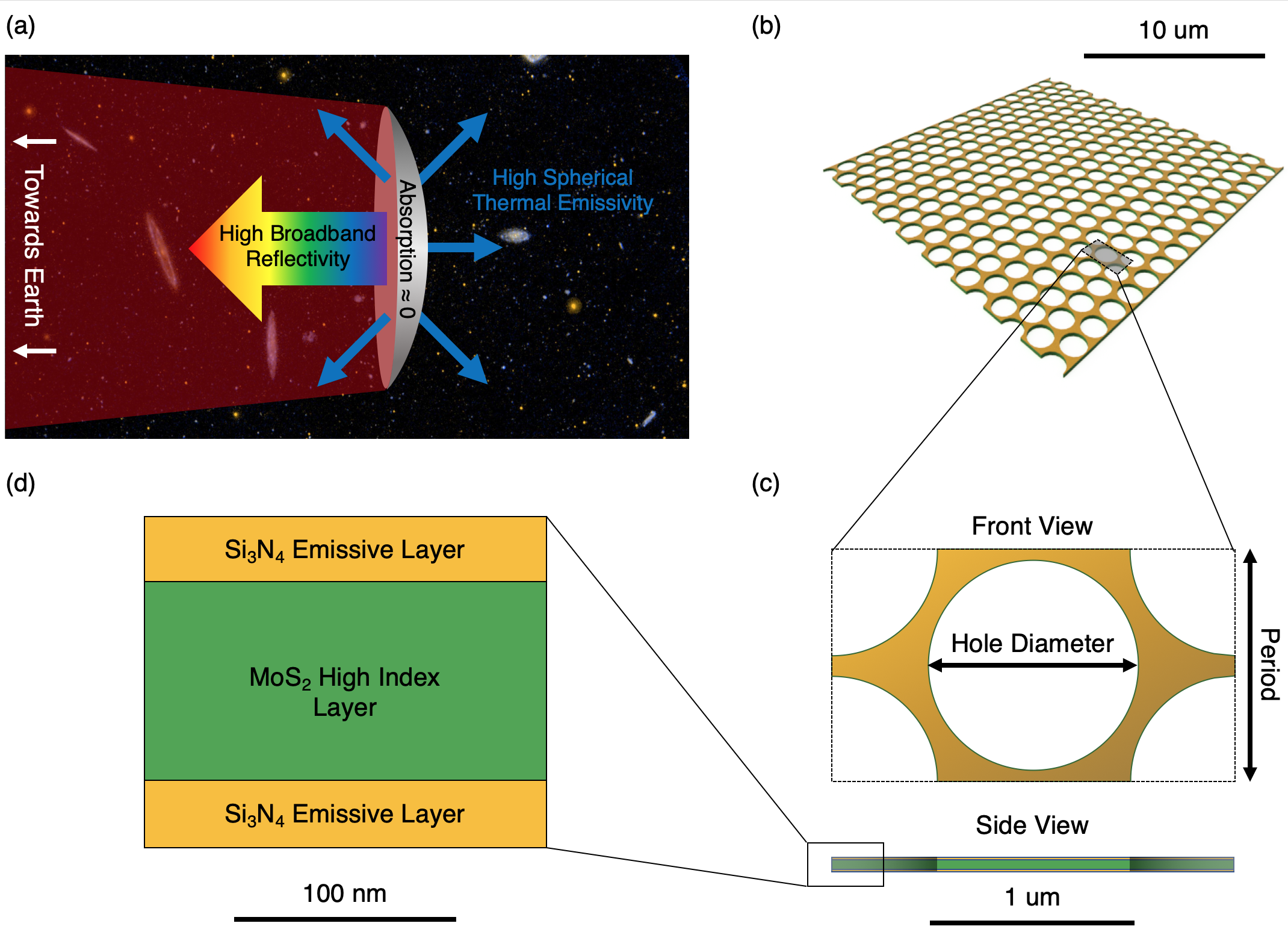}
\caption{Starshot sail design. (a) Schematic diagram demonstrating relevant optical considerations for an accelerating lightsail (b) Section of sail with approximate hole diameter to period ratio of 90\% (c) Front and side view of single design period. (d) Enlarged view of multilayer structure. Yellow regions represent \ch{Si3N4} while green regions represent \ch{MoS2}.}
\label{figure1}
\end{figure}

\newpage With this in mind, a central challenge in determining the survivability of the sail lies in mapping the trade-off between the sail being minimally absorptive across the Doppler-shifted laser wavelengths and having high emissivity at longer thermal wavelengths, given the need to maximize reflectivity in order to minimize the acceleration distance. Motivated by this, we developed a multilayer 2D photonic crystal slab-based geometry that features molybdenum disulfide (\ch{MoS2}) and silicon nitride (\ch{Si3N4}) as its key constituent materials. Figure~\ref{figure1}a illustrates the general optical considerations of lightsail design, while Figures \ref{figure1}b and \ref{figure1}c show the hole diameter and period parameters we investigated in our periodic concept sails. Figure~\ref{figure1}d details the thickness parameters of the layered structure of our design, which is composed of two emissive \ch{Si3N4} outer layers surrounding a \ch{MoS2} high-refractive index reflective core. 

Previous work has discussed the possible use of MoS\textsubscript{2} for laser-driven lightsails\cite{Atwater2018}, but its exploration in plausible sail designs is still unexplored. Here we choose to investigate designs employing \ch{MoS2} for three key reasons. The first is its high refractive index in the Doppler-shifted laser band (roughly 1.2-1.47 um, to minimize atmospheric absorption by the Earth-based laser array), ranging from n = 3.73 - 3.66, which makes it a desirable reflective material. Second, monolayer \ch{MoS2} samples have been shown to have zero absorption in the laser bandwidth within measurement limits\cite{Ermolaev2020}. Third, large area monolayer samples have been fabricated successfully, a significant step toward future lightsail-scale films\cite{Dumcenco2015,Li2020,Nie_2017}.

\ch{Si3N4} has also been recently investigated for lightsails\cite{Myilswamy2020,doi:10.1021/acsphotonics.0c00768,Gao2020} and remains a well-qualified lightsail candidate material due to its mature fabricability, low density, and  high decomposition temperatures\cite{nishi2000handbook,haynes2014crc}. 
In addition to these considerations, our primary motivation for using this material is its desirable thermal emissivity at wavelengths longer than the Doppler-shifted laser band, which will be discussed in detail below\cite{Kischkat2012}.
As used in our sail design, \ch{Si3N4} acts as the primary radiative cooler, preventing thermal failure during acceleration. We emphasize that in our design we do not assume the presence of an additional layer or layers that provide non-zero emissivity. Rather, the combined sail design is meant to be holistic and comprehensive.

Due to the stringent mass constraints of Starshot, our design involves a 2D photonic crystal with a close-packed array of large-radius holes with respect to the lattice constant of the structure. In addition to the mass reduction benefits, the photonic design provides reflective enhancement through coupling to broadband guided modes, building on conventional 2D photonic crystal slab theory\cite{Fan2002}. However, the extreme performance required of the lightsail necessitates designs typically not employed by conventional photonic crystal reflectors. Specifically, to our knowledge, previous conventional photonic crystal reflectors in the literature have not been severely mass constrained. Structurally, an important advantage of our proposed design is that it is fully connected and requires no additional substrate to function as a standalone sail\cite{doi:10.1021/acsphotonics.0c00768}. In practice however, we expect it will be beneficial to use a large corrugated support backbone to allow the sail to withstand the extreme forces it will undergo as it accelerates\cite{Davami2015}. This structure would provide macroscale sail curvature to increase stability and mechanical robustness\cite{campbell2021relativistic} while additionally limiting crack propagation in our proposed designs due to patterned hole-induced stress concentrations. 

\begin{figure}[h]
\includegraphics[width=16cm]{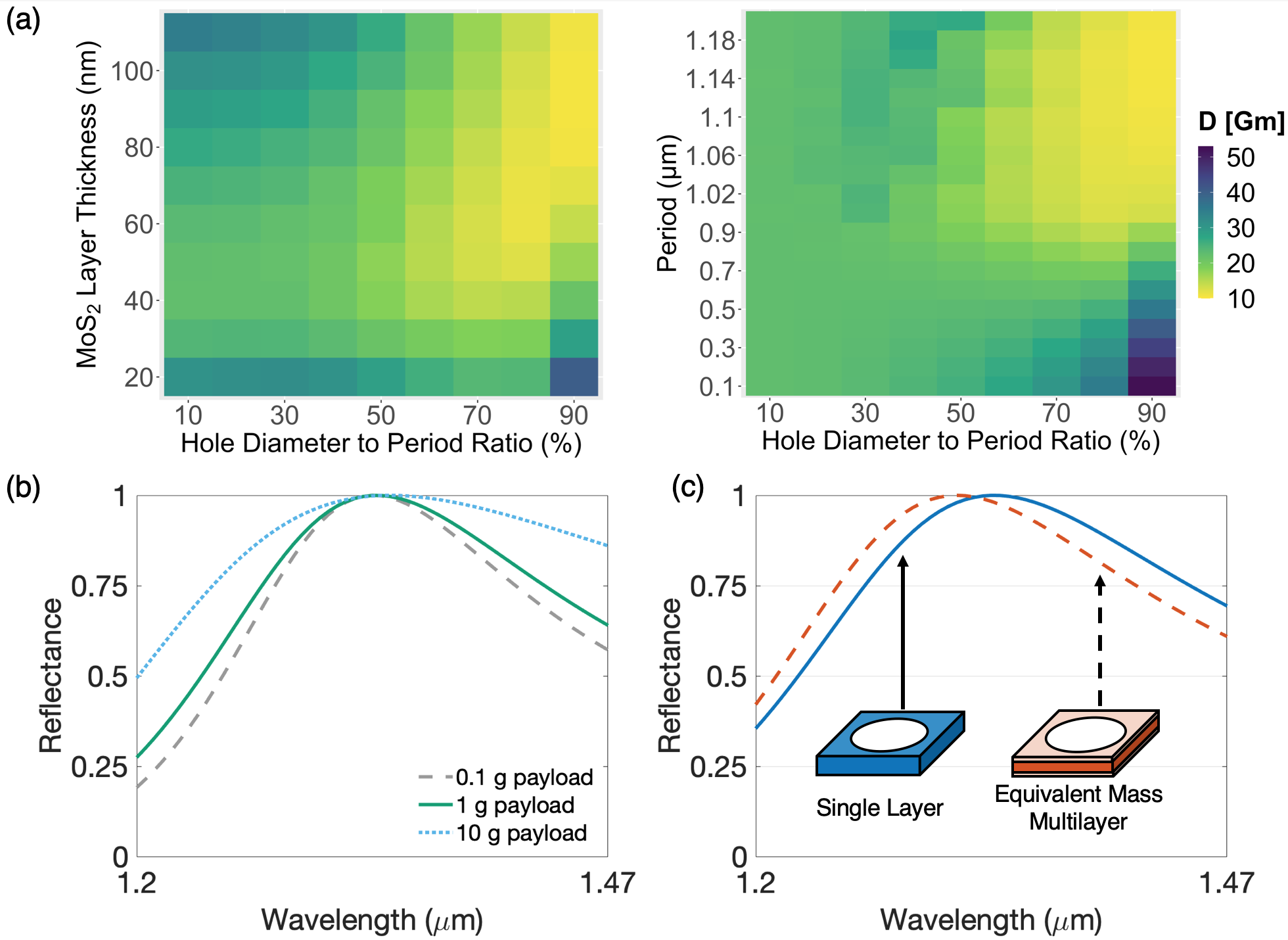}
\caption{Reflective properties of multilayer photonic sail structures. (a) Color maps of minimum acceleration distance designs as a function of most sensitive geometric parameters. (b) Dependence of sail reflection spectra on payload mass over the Doppler-shifted laser wavelength range. Lower mass payloads reward reductions in the sail mass more than increases in the sail's integrated reflection spectra. (c) Demonstration of desirable spectral perturbation due to the addition of lower index, high emissivity \ch{Si3N4} layers. Insets provide schematic diagrams of sail designs corresponding to plot colors.}
\label{figure2}
\end{figure}

To merit sail designs, we assumed typical values for the Starshot Initiative: a uniform I = 10 $GW/m^2$ laser irradiance on the sail, a laser output wavelength of $\lambda = 1.2~\mu m$, a 10 $m^2$ sail area, and a $\sim$1 $g$ chip payload mass unless otherwise stated. The acceleration distance figure of merit, as defined by \citeauthor{doi:10.1021/acsphotonics.0c00768}\cite{doi:10.1021/acsphotonics.0c00768} is:
\begin{equation}\label{accel_FOM}
D=\frac{c^3}{2I}(\rho_{payload}+\rho_{sail})\int_{0}^{0.2}\frac{h(\beta)}{R(\lambda(\beta))}d\beta
\end{equation}
where $h(\beta) = \frac{\beta}{(1-\beta)^{2}\sqrt{1-\beta^{2}}}$, I is the laser irradiance in $\frac{W}{m^2}$, $\rho$ is areal density in $\frac{kg}{m^2}$ (discussed more below), $\beta$ is the unitless relative velocity (to the speed of light), and $R(\lambda(\beta))$ is the spectral reflectance over the Doppler-shifted laser band.

Optimizing the period and hole diameter of the patterned holes allows for high transmission and reflection bands of varying spectral bandwidth\cite{Fan2002}. Figures \ref{figure2}a and b demonstrate the dependence of acceleration distance on key design parameters in our sail design space. Each color map represents a two dimensional slice of the five dimensional design space composed of the period/lattice constant, the hole diameter-to-period ratio, and the thicknesses of each of the three layers. The tile colors represent the minimum acceleration distance design possible for the parameter values specified on the axis (this minimum is achieved by changing the unseen parameters to their optimal values for the given tile). Our optimal acceleration distance merited design has a period of 1.16~$\mu$m, a hole period to diameter ratio of 90\%, 5~nm thick emissive \ch{Si3N4} layers, and an 90~nm thick \ch{MoS2} reflective core, placing it in a regime of very low values of thickness relative to the lattice constant, $<$ 0.1$a$. Thinning of the high-index core maintains access to broad-band Fabry-Perot-like reflection modes at normal incidence with the added effect of minimizing the overall sail mass. This demonstrates the broad range of possible acceleration distance values that our design space encompasses. 

While the actual mass of the payload chip has not been determined yet, it is important to understand the relative effect of the payload mass on our optimal reflective design. Note that payload mass can be converted to the areal density value shown in (\ref{accel_FOM}) easily by dividing its mass by sail area: $\rho_{payload} = m_{payload}/A_{sail}$. Figure \ref{figure2}c plots the laser band reflection spectra of the lowest acceleration distance design for three given payload weights, demonstrating that as the payload mass increases, reducing the sail's mass is rewarded less than increasing its integrated reflectance. This means that sail mass becomes a stronger consideration when the payload mass is small. A further analysis showing the minimum acceleration distance vs. payload mass can be found in the Supporting Information, which is corroborated by results in \citeauthor{doi:10.1021/acsphotonics.0c00768}\cite{doi:10.1021/acsphotonics.0c00768}

The \ch{Si3N4} layers have primarily been introduced to enhance thermal emissivity; however, these outer layers can also have the effect of shifting the peak of the sail reflection spectra to lower wavelengths compared to that of single-layer \ch{MoS2}-only designs, as shown in Figure \ref{figure2}c. This shift fortuitously results in an improvement in the acceleration distance figure of merit. Note that since the designs shown in Figure \ref{figure2}c have identical masses; this effect is attributable to the change in the refractive index profile alone.

The optimization procedure described thus far yields an optimal reflective design in this sample set of 10.6~Gm with a 1~g payload, comparable in performance to the best reported numbers in the literature\cite{doi:10.1021/acsphotonics.0c00768}. Importantly, this design does not require a connecting support structure and all mass required for acceleration and cooling is accounted for in this figure of merit value. Additional structural stability may be provided by a 1-g mechanical backbone structure, giving a value of 15.2~Gm. Furthermore, the topology of this design is not computationally optimized, and optimization could yield still lower acceleration distances. While this design is competitive with others shown previously\cite{doi:10.1021/acsphotonics.0c00768,Salary2020} on acceleration distance metrics alone, lightsails with realistic additional thermal considerations require 54\% larger acceleration distances, as we show next.

\begin{figure}[h]
\includegraphics[width=16cm]{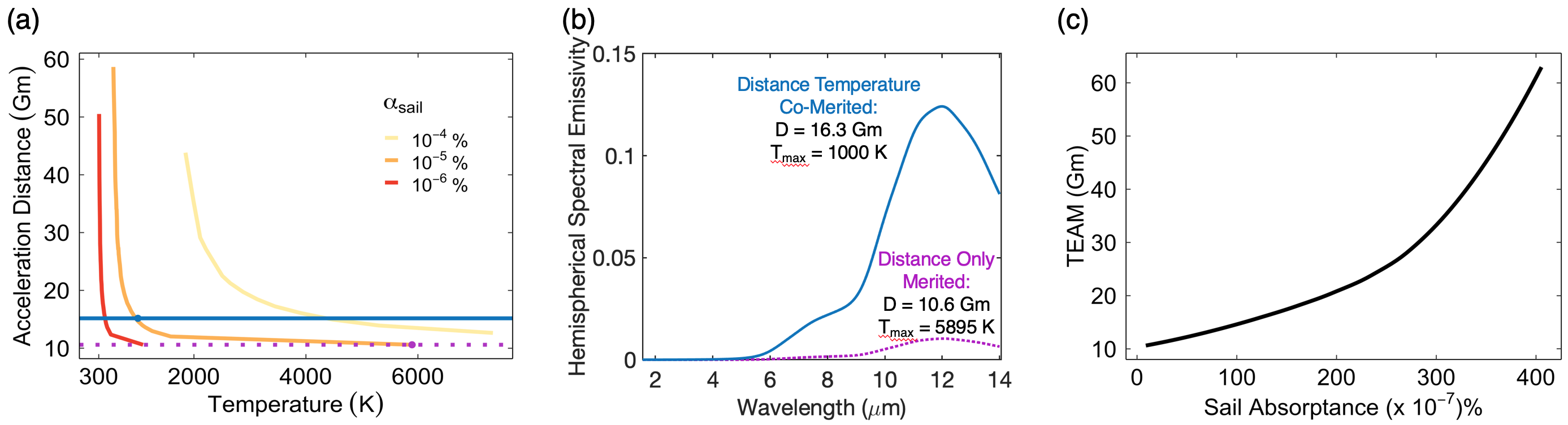}
\caption{Emissive properties of multilayer photonic sail structures. (a) Minimum acceleration distance as a function of temperature for assumed values of sail absorptivity. Points and horizontal lines demonstrate how selected designs in (b) change temperature as assumed absorptivity is altered. (b) Emissivity comparison of simple acceleration merited design vs. acceleration temperature co-merited design. Acceleration distance and maximum temperature of each design are also shown. (c) Plot showing the thermally endurable acceleration minimum (TEAM) distance value of our design space as a function of assumed lightsail absorptivity. TEAM is the thermally-constrained minimal acceleration distance design.}
\label{figure3}
\end{figure}

Maintaining the sail's integrity as it accelerates is a fundamental consideration in design, more important than any acceleration distance or reflection-driven performance metric. Though the sail will exhibit extremely low absorbance, its temperature will nevertheless increase due to the high incident laser photon flux, and its interaction with the interstellar medium at relativistic speeds could cause further heating\cite{Hoang_2017}. Unfortunately, the sail's component materials will likely become more absorptive as their temperature rises, causing thermal runaway effects and increasing the probability of sail mechanical failure due to material degradation. The radiative cooling characteristics of the sail are therefore extremely important. As a thermal limit, we have adopted the ultra-high vacuum (UHV) sublimation temperature $T_{sublimation}$ of the sail materials, which is the point at which the sail would begin to spontaneously evaporate and/or decompose. Note that, since the UHV sublimation temperature is less than the melting temperature, which has been selected as the thermal limit in other recent lightsail studies, this represents a relatively conservative design decision. In our case, we adopted $T_{limit}=T_{sublimation,\ch{MoS2}} \sim 1000$~K or the lower UHV sublimation point of the two materials used (see Supporting Information).

To proceed, we implicitly calculated the maximum temperature $T_{max}$ reached by each sail in our space of over $3\times10^5$ designs using the following equation:
\begin{equation}\label{temp_FOM}
P_{laser}\alpha_{sail}=2A_{sail}\int_{a}^{b}\frac{c_{1}}{\lambda^{5}}\cdot\frac{\epsilon_{sail} (\lambda)}{e^{\frac{c_{2}}{\lambda T_{max}}}-1}d\lambda
\end{equation}
where $P_{laser}$ is the output laser power (100 GW), $\alpha$ is the assumed normalized sail absorption, $A_{sail}$ is the area of a single side of the sail, the factor of two accounts for the presence of \ch{Si3N4} on both sides of the sail, $c_{1}=2\pi hc^{2}$, $c_{2}=hc/k_{b}$, $h$ is Planck's constant, $c$ is the speed of light, $k_b$ is Boltzmann's constant, and $\epsilon_{sail}$ is the spectral sail emissivity. 

Note that the arbitrary choice of total sail absorption in (\ref{temp_FOM}) is due to the present lack of sufficiently sensitive material extinction coefficient data in the Doppler-shifted laser wavelength band. This highlights the need for ultra-high sensitivity measurements of material absorption characteristics using techniques such as photothermal deflection spectroscopy\cite{Boccara1981} or photocurrent spectroscopy\cite{Keevers1995} in order to qualify lightsail materials. The values we used in Figure \ref{figure3}a demonstrate that material absorption must be miniscule in order for sails to perform comparably to others in literature. 

We now define a new and final composite figure of merit for our sail design space, which we call the thermally endurable acceleration minimum (TEAM) distance value. The TEAM distance value for a design space is that for which the acceleration distance D is minimized, among the alternatives for which $T_{max}<T_{limit}$. Likewise, the TEAM sail design is the sail configuration that results in the TEAM distance value. Minimizing TEAM distance is desirable, but we emphasize that this is not a sail \emph{metric} per se; rather, it is a single summary value that can be easily reported to compare design approaches and sail datasets, as opposed to individual sails. For a constant set of laser parameters and a given set of sail architectures, this final result is dependent on two quantities: the previously assumed maximum allowable sail temperature set by the UHV sublimation limit, and the previously assumed total absorptance of the sail. 

While conventional photonic crystal slabs have desirable reflective properties from an accelerative standpoint, their thermal radiant exitance is highly dependent on the intrinsic spectral emissivity of their component materials. In the case of lightsails, a strong trade-off exists between having sufficient emissivity for heat dissipation and minimizing the acceleration distance. In particular, for a given sail diameter, while acceleration distance is generally negatively impacted by mass increases, thermal emissivity is generally rewarded by increasing the amount of emissive material per unit area. In analyzing our sail designs, we assume total sail absorption values ranging from of $10^{-4}$\% to $10^{-6}$\% and show in Figure \ref{figure3}a the relationship between acceleration distance and operating temperature for three of these absorption values.

Using our analysis framework, we demonstrate a TEAM distance value of 16.3~Gm, a 5.7~Gm accelerative penalty to prevent decomposition due to sublimation of sulfur out of the \ch{MoS2} in the sail by limiting the sail's temperature to T$_{max}=1000$~K. Addition of a 1~g mechanical backbone results in a larger TEAM value of 21.3~Gm. The reduced maximum temperature of the design that achieves the TEAM value is due to its smaller hole radius and thicker emissive \ch{Si3N4} layers, which imply that more material is present to radiate away excess energy relative to the 10.6~Gm acceleration distance design. This can be seen in the comparison of the spectral hemispherical emissivity values of the TEAM design vs. the baseline simple acceleration distance merited design in Figure \ref{figure3}b. The TEAM design an approximately $12\times$ higher peak emissivity value due to the presence of more \ch{Si3N4} in the photonic crystal design. Alternatively, one can analyze the effect on the TEAM value as a function of overall sail absorption in the laser bandwidth. As can be seen in Figure \ref{figure3}c, when $\alpha$ increases, the TEAM distance value will also increase, placing firm bounds on material absorption of the incident laser light in order to achieve a certain acceleration distance.

\begin{figure}[h]
\includegraphics[width=16cm]{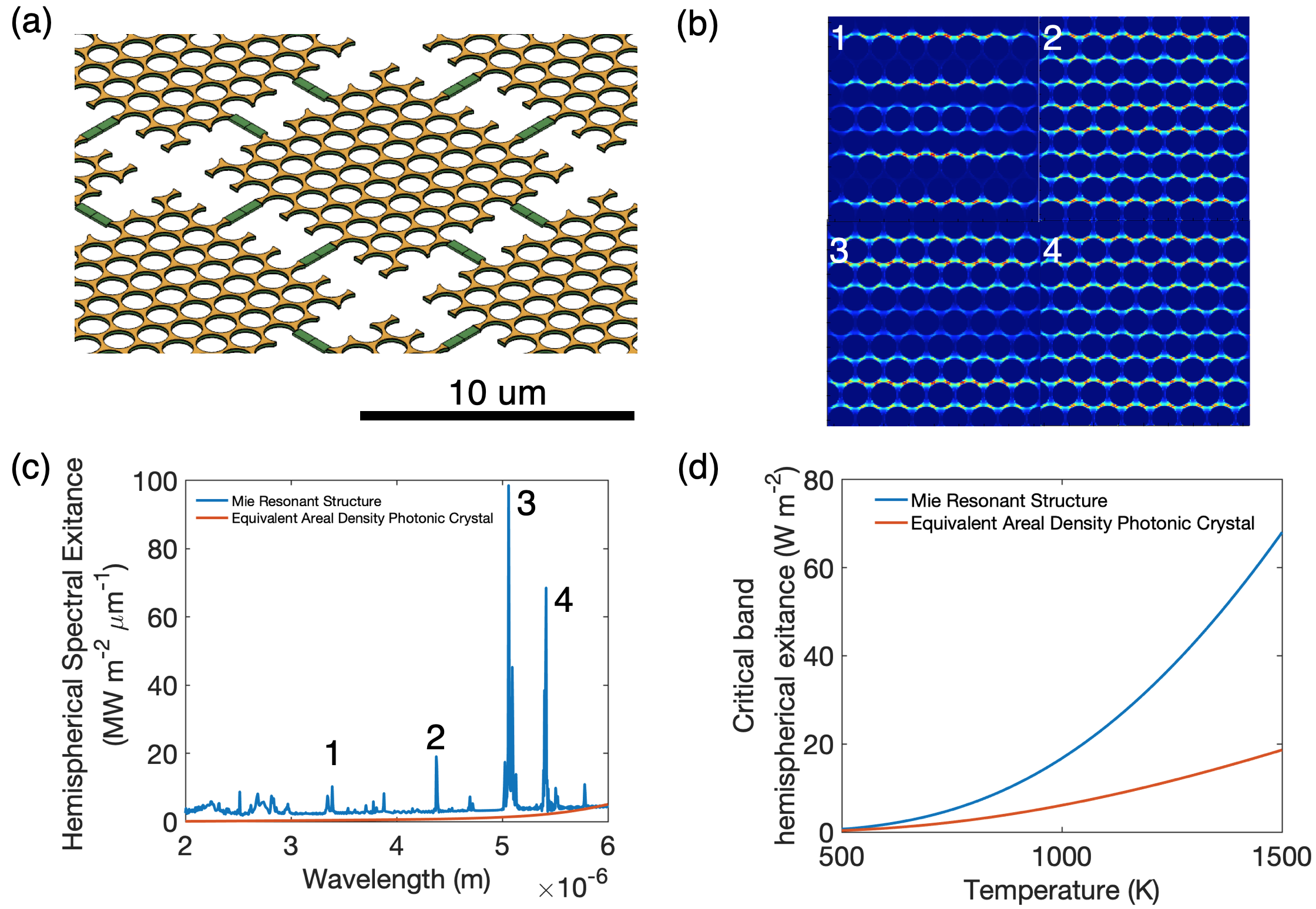}
\caption{(a) Mie resonant enhancement schematic diagram illustrating the proposed multiscale Mie resonator design. The green strips show the possibility of scaffolds used to support the Mie resonator structure. (b) Spatial absorption profile of Mie resonances (c) Hemispherical exitance of Mie resonant structure vs. conventional photonic crystal structure, calculated at a temperature of 1000~K, demonstrating a pathway for possible emissivity enhancement. Labelled wavelength peaks correspond to spatial profiles in (b). (d) Demonstration of spectrally integrated hemispherical exitance enhancement as a function of temperature.}
\label{figure4}
\end{figure}

%\section{Emissivity enhancement through Mie Structure Engineering}

Previously developed photonic designs for laser lightsails have only employed single-scale photonic features to increase sail reflectivity and analyzed any thermal characteristics in that context. However, a multiscale segmented design architecture has the possibility to yield additional thermal benefits without incurring a mass penalty. To this end, we propose a multiscale Mie resonant structure-based approach to emissivity enhancement that employs thermal wavelength-scale Mie resonators patterned with laser wavelength-scale photonic crystal features, shown schematically in Figure \ref{figure4}a. These multiscale photonic structures yield a large number of resonant peaks over desired peak thermal wavelengths that collectively enhance the total hemispherical emittance of the sail. This is reminiscient of nanophotonic light trapping approaches to solar absorption, where numerous peaks contribute to a large scale enhancement of absorption\cite{Yu2010,Yu2012}; here we apply these light trapping techniques for emissive gains. Our design is composed of two 30~nm emissive \ch{Si3N4} layers surrounding a 90~nm \ch{MoS2} core. The patterned photonic crystal has a period of 1.14~$\mu$m, with a 90\% hole diameter to period ratio. The Mie structure is 8.55~$\mu$m by 8.55~$\mu$m and the overall period is 10~$\mu$m. While not a direct comparison due to geometric differences (ie, a sail with the small-scale photonic structures but not the large-scale islands), a continuous design in the space investigated with similar areal density per layer to the Mie structure reveals the continuous design has acceleration distance of 24 Gm, while the Mie resonant design has an acceleration distance of 16.7 Gm. Thus the Mie resonant design has a 7.4~Gm shorter acceleration distance or a 43\% decrease as compared to the equivalent continuous design. Because the areal density of \ch{Si3N4} between the compared designs is nearly the same, longer wavelength thermal emissive features will be maintained between the two desigbs, meaning the additional emissivity features in the critical band from 2-6~$\mu$m are key to enable lower overall temperatures. The structure can be connected by a series of thin scaffolds while maintaining the presence of resonant modes. If further mechanical robustness is desired, a mechanical backbone could be added (another approach for further structural stability is also investigated in the Supporting Information). 

The spatial profiles of four resonant modes supported by the multiscale Mie-resonant structure, are shown in Figure \ref{figure4}b, corresponding to four modes in the 2 - 6~$\mu$m band shown in Figure \ref{figure4}c. This wavelength band is critical for sail heat management due to the blackbody peak position at temperatures from 500 - 1000K, as determined by Wien's law. Increases to sail emissivity in this band will more strongly reduce overall sail temperatures in comparison to emissivity increases at longer wavelengths. The enhancement of in-band hemispherical exitance at a given temperature is demonstrated in Figure \ref{figure4}d, showing that at the previously suggested thermal limit of 1000~K, the islanded design has over $2.75\times$ greater hemispherical exitance, with as much as $3.6\times$ the hemispherical exitance at 1500~K. This showcases the utility of the multiscale Mie-resonant structures in sail thermal regulation.
 
%\section{Conclusion}
In conclusion, we have demonstrated holistically viable multilayer 2D photonic reflector designs for laser-driven lightsails that are able to accelerate to one fifth the speed of light over distances comparable to, and in some cases even exceeding, designs reported previously. We emphasize that our designs represent the entire sail structure and do not require additional backing material for emissivity enhancement, allowing for accurate modeling of payload-driven performance. To analyze such relativistic lightsail designs, we further proposed an analysis framework that judges sail designs according to both their acceleration distance and peak temperature. We then proposed the thermally endurable acceleration minimum (TEAM) distance value as a summary statistic to determine the fastest-accelerating thermally-stable sail design of a design set. This value is easily reportable and will allow future engineers to compare their design sets, represented by a variety of materials, nanoscale geometries, etc. Finally, we introduced a multiscale sail design employing thermal-wavelength-scale Mie-resonant features to act as a means of enhancing the mid-infrared emissivity of lightsails while preserving their underlying acceleration distance characteristics. Although the goals of Breakthrough Starshot impose stringent constraints, our multi-scale photonic designs highlight intriguing optical capabilities that ultralight, nearly-massless photonic structures can enable over an ultrabroadband wavelength range. This in turn heralds the possibility of new classes of ultralight photonic structures that can perform as well as conventional photonic structures, and thus enable new capabilities in mass-constrained applications.

\section{Methods}
We performed reflective simulations using the S$^4$ RCWA solver \cite{Liu20122233} and Lumerical FDTD Solver. Simulations were performed over a Doppler-shifted wavelength band from 1.2-1.47~$\mu$m for reflection. We performed absorptivity simulation in S$^4$ from 1.55-14~$\mu$m, which is detailed in the Supporting Information. Data was analyzed and merited using MATLAB R2020a and RStudio.  

\begin{acknowledgement}

This work was supported by the Breakthrough Initiatives, a division of the Breakthrough Prize Foundation. J.B. is supported by a National Science Foundation Graduate Research Fellowship under grants DGE-1650605 and DGE-2034835. It was also funded in part by a National Science Foundation CAREER award under grant CBET-1845933.
\end{acknowledgement}

\bibliography{revisedmanuscript051421}

\end{document}